\documentclass[%
 reprint,
superscriptaddress,
amsmath,amssymb,
 aps,
 prl,
floatfix,
]{revtex4-2}

\usepackage{physics}
\usepackage{color}
\usepackage{graphicx}
\usepackage{dcolumn}
\usepackage{bm}
\usepackage{calc}
\usepackage{verbatim}
\usepackage{float}
\usepackage{svg}

\usepackage[mode=buildnew]{standalone}
\begin{document}

\title{Phase-sensitive avalanche quantum sensing of sub-shot-noise fields}

\author{Nikolai D. Klimkin}
\email{Nikolay.Klimkin@mbi-berlin.de}
 \affiliation{Max Born Institute, Max-Born Stra{\ss}e 2A, 12489 Berlin, Germany}
\author{Misha Ivanov}%
\affiliation{Max Born Institute, Max-Born Stra{\ss}e 2A, 12489 Berlin, Germany}
\affiliation{Institute of Physics, Humboldt University Berlin, 12489 Berlin, Germany}
\affiliation{Technion – Israel Institute of Technology, 3200003 Haifa, Israel}

\begin{abstract}
    Avalanche-based detection of small perturbations is commonplace in precision measurement devices from Geiger counters to single-photon avalanche detectors. Here, we expand this principle to sensing of light waves below the quantum shot noise limit, demonstrating numerically how atomic clouds trapped in photonic cavities can exhibit non-perturbative sensitivity to changes in the cavity field when the cavity mode is tuned to a parity-prohibited transition. We find that, when pumped by a continuous-wave laser, the atomic cloud creates emissions which amplify the initially sub-shot-noise fluctuation by two orders of magnitude. Crucially, this amplification is broadly tunable with respect to frequency and independent of atomic energy structure. Even more strikingly, our amplification mechanism preserves the phase imparted on the atoms by the original ultra-weak light wave, marking a qualitative improvement over conventional protocols. Our finding has implications for both quantum state characterization and detection of weak classical signals.
\end{abstract}

\maketitle

Avalanche detection is a principle applied in a wide range of devices to achieve an exponential boost in the detector's sensitivity to external perturbation. Suppose a perturbation is too small to disrupt the detector in its thermodynamic equilibrium. One typical example of such a perturbation is the flyby of a single particle of radiation. When this is the case, a device is prepared in a metastable state which is engineered to collapse at a slight disturbance. In the case of ionizing radiation, the first detector devices are cloud chambers where the incident particle triggers runaway condensation of superheated vapor, as well as Geiger-Müller tubes forming the basis of Geiger counters where an ionizing particle triggers a runaway electric discharge. In modern detection setups, single-photon avalanche detectors employ the same principle of runaway ionization, leveraging the metastable state of a semiconductor at high voltage to provide highly sensitive detection of isolated photons. In every one of these concepts listed, the external perturbation constitutes a purely classical change in external force. %Such a measurement allows one to retrieve its presence and time of arrival, but not the quantum phase such an excitation may have carried. 

At the same time, reconstructing the vacuum fluctuations of the quantum electromagnetic field is also proving to be of interest, pioneered in a novel line of work~\cite{riek2015direct} demonstrating their retrieval by electro-optical sampling. Here, we apply the principle of avalanche detection to fluctuations of quantum electromagnetic vacuum, characterized by both an amplitude and a quantum phase. We demonstrate that atomic clouds in properly detuned optical cavities can form a quantum avalanche detector when pumped by a classical laser wave. Like a typical avalanche detector, this setup uses the principle of a collapsible metastable state to transform an atomic-scale perturbation into macroscopically-detectable change in observables. However, in addition to that, a quantum avalanche detector does not destroy phase information, allowing one to reconstruct the precise quantum state of the cavity.

Atomic ensembles whose collective emission bursts are triggered by vacuum fluctuations have been investigated extensively in the context of Dicke superradiance~\cite{dicke1954coherence}. In a conventional superradiant setup, a fully inverted ensemble of atoms begins decaying at a classical rate of $1$, accelerating by a factor of $N$ as the atoms transition into a correlated state. In our case, the ensemble is placed in a cavity, with the cavity mode spectrum centered around a parity-forbidden transition. As such, our classical emission rate is close to zero. Any quantum fluctuation in these modes becomes self-amplifying, triggering an emission whose back-action on the emitters unlocks ever more emission. As investigated in \cite{klimkin2025spontaneous}, this causes entanglement to develop between the quantum field and the emitter states, with the emitter dipole being locked to the quantum vacuum fluctuations, which in their turn are amplified by the dipole emission. Here, we propose an application of the described mechanism, demonstrating how this self-amplification can lead to a drastic improvement to the accuracy of quantum sensing techniques.

Demonstrated in experiment to be a promising single-photon source~\cite{muller2015coherent}, emission into detuned photonic modes poses formidable theoretical challenges. 
The conventional approaches, which view the entire mode continuum through an effective perturbative description \cite{pizzi2023light, yi2025generation}, or use a non-perturbative description but constrain the entire spectrum to a few effective modes \cite{tziperman2023quantum}, are insufficient when considering large arrays of strongly driven emitters strongly coupled to photonic modes. Here, one faces the challenges of (i) describing strongly correlated many-body dynamics of quantum emitters developing through the coupling to a common light mode~\cite{muller2025genuine}, (ii) accounting for possible symmetry breaking by strong quantum fluctuations in emitters~\cite{stitely2023quantum}, all in conjunction with (iii) arbitrarily large quantum states of light generated
via nonlinear optical response to a strong external driving field. The broad emission spectra and the large numbers of the emitted photons typically force one to use a Markov-regime description of the emitters' interaction with the quantum vacuum. The Markov regime would rule out any emitter-mode entanglement, and by extension generation of nontrivial states of light. Our approach (see Supplementary information for details) overcomes these challenges.

\begin{figure}[b]
    \centering
    \includegraphics[width=\linewidth]{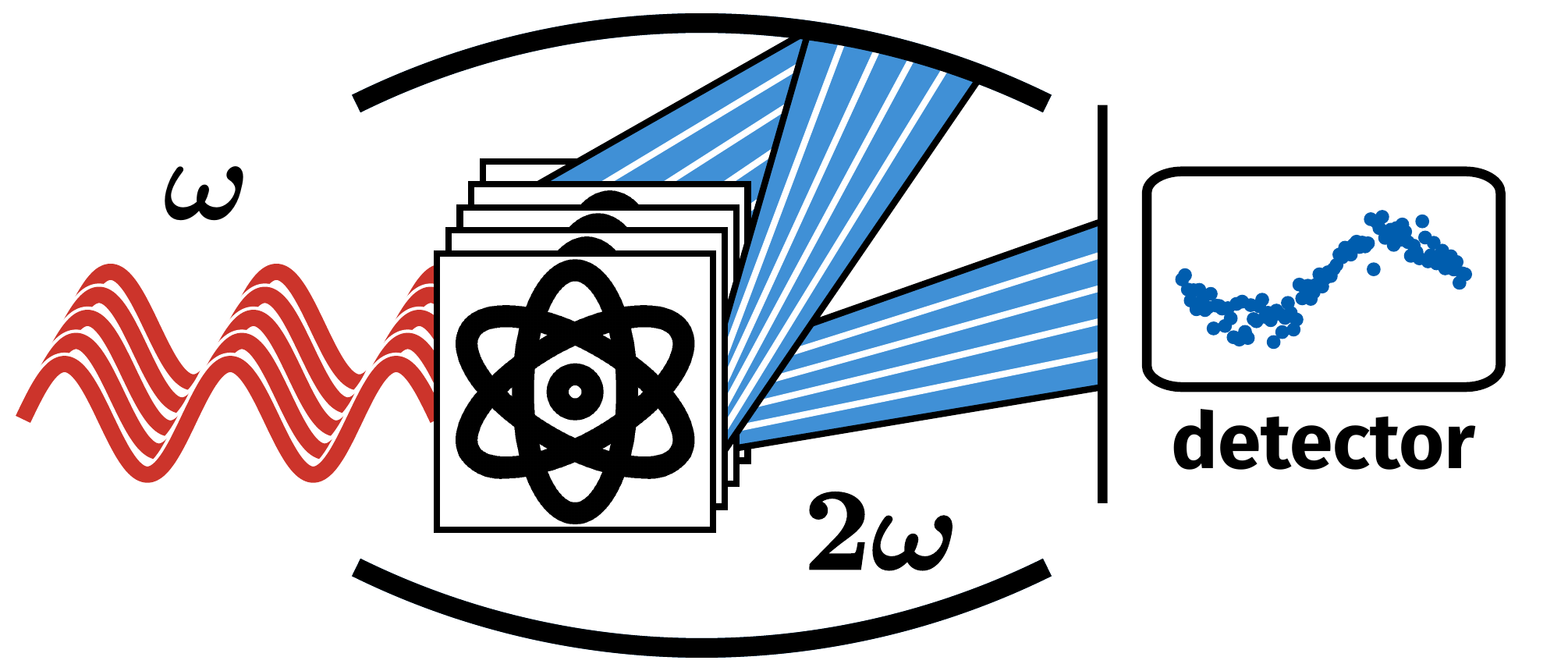}
    \caption{(color online) Basic design of our setup. The classical radiation (red), resonant in our case with the atomic transitions, drives ultrafast currents in all atoms independently. This causes them to emit frequency-converted photons (blue) entangled with the emitters. The electromagnetic radiation confinement in a medium causes these photons (blue beams with white lines) to linger, allowing them to either be absorbed by the same or another atom, or leave the confining medium. In the latter case, they can be observed by a classical detector positioned outside (right).}
    \label{fig:fig1}
\end{figure}

Fig.~\ref{fig:fig1} depicts our system of many $N_e\gg1$ atoms (here, $N_e=60$) inside a cavity. By having a finite bandwidth, this cavity incorporates a unitary way for the photons to separate themselves from the emitters. A similar effect can be achieved in e.g. a waveguide or a photonic crystal with a sufficiently low group velocity~\cite{busch1999liquid, busch2000radiating}. The atoms are modeled as two-level systems with a resonance frequency $\omega_0$, subjected to a classical driving force. The classical driving field %shown in Fig.~\ref{fig:multiplot-2h}(a) 
is off-resonant with the atomic transition: $\omega = 0.625\omega_0$. It has a flat-top time profile and a peak Rabi frequency of $0.1 \omega_0$. This classical field's carrier envelope phase (CEP) is assumed not to be stabilized between measurements. As such, all quantities presented in this work are averaged with respect to the field CEP. The cavity's resonance is centered at $\epsilon = 1.25\omega_0$, being off-resonant from the atomic transition while resonant with the driver's second harmonic. As such, both the driving field and the atomic resonance are significantly off-resonant from the atomic transition, proving our protocol to be independent from the atomic structure. As second harmonic generation is forbidden in a centrally symmetric medium, the cavity suppresses uncorrelated emission by individual atoms.

\begin{figure}[b]
    \centering
    \includegraphics[width=\linewidth]{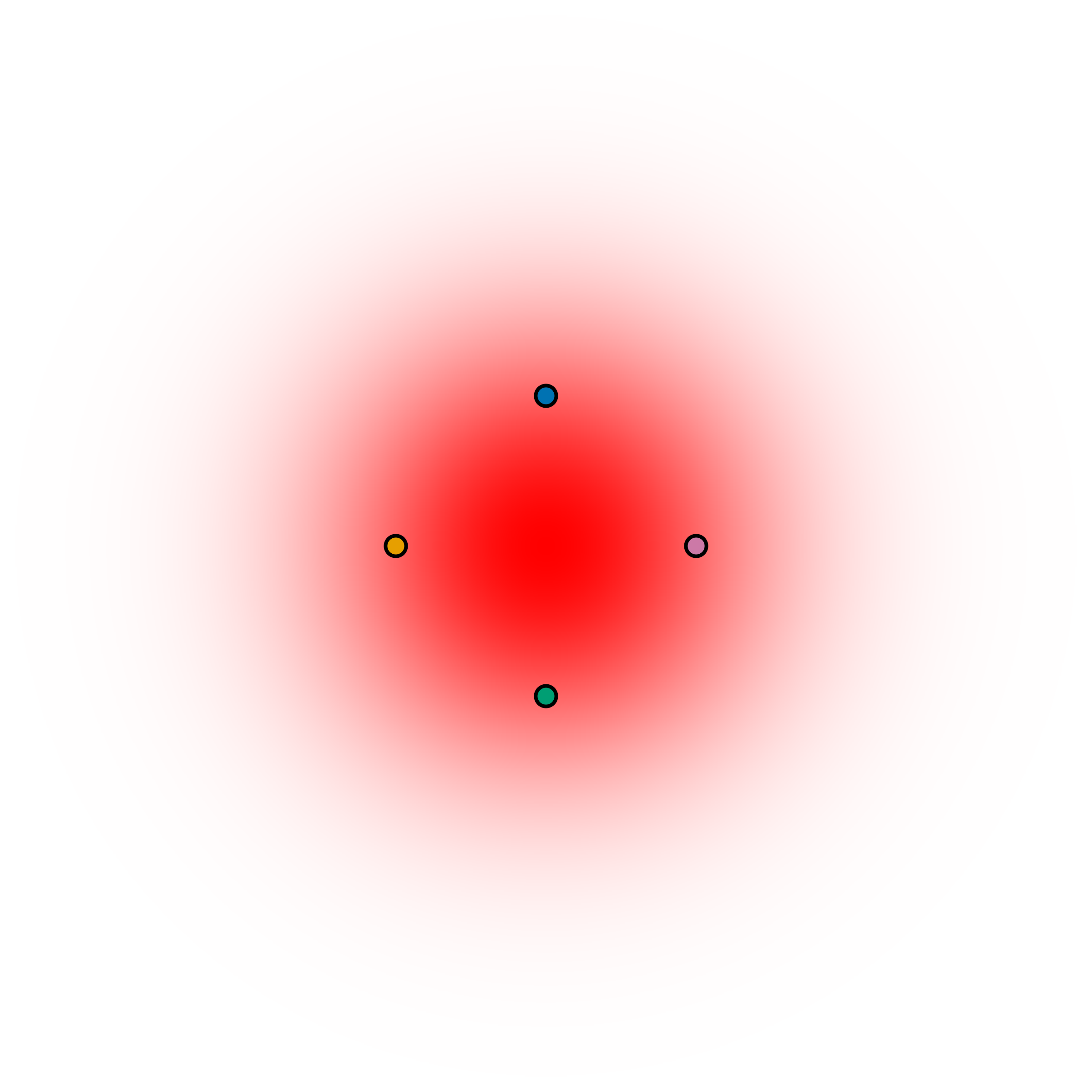}
    \caption{Vacuum Husimi function overlaid with the coherent displacement amplitudes $\alpha_p$ used in this work as perturbations.}
    \label{fig:pert}
\end{figure}

To demonstrate avalanche sensing in this setup, we then introduce a non-trivial initial state of the cavity's photonic field. Specifically, instead of vacuum, the initial state of the cavity is set to a perturbing coherent state $\ket{\alpha_p}$. We choose $\alpha_p = |\alpha_p| e^{i\varphi_p}$ for varying $\varphi_p$, and constant $|\alpha_p|^2 = 0.5$. Fig.~\ref{fig:pert} demonstrates the magnitude of this displacement compared to the ordinary vacuum fluctuations inherent to a photonic mode. This sub-shot noise displacement is nearly impossible to detect by a macroscopic device affected by a thermal noise which is at least comparable in magnitude to vacuum fluctuations. Below, we demonstrate amplification of this initially small field by more than an order of magnitude, relaxing the requirements on such a device.

\begin{figure}[b]
    \centering
    \includegraphics[width=\linewidth]{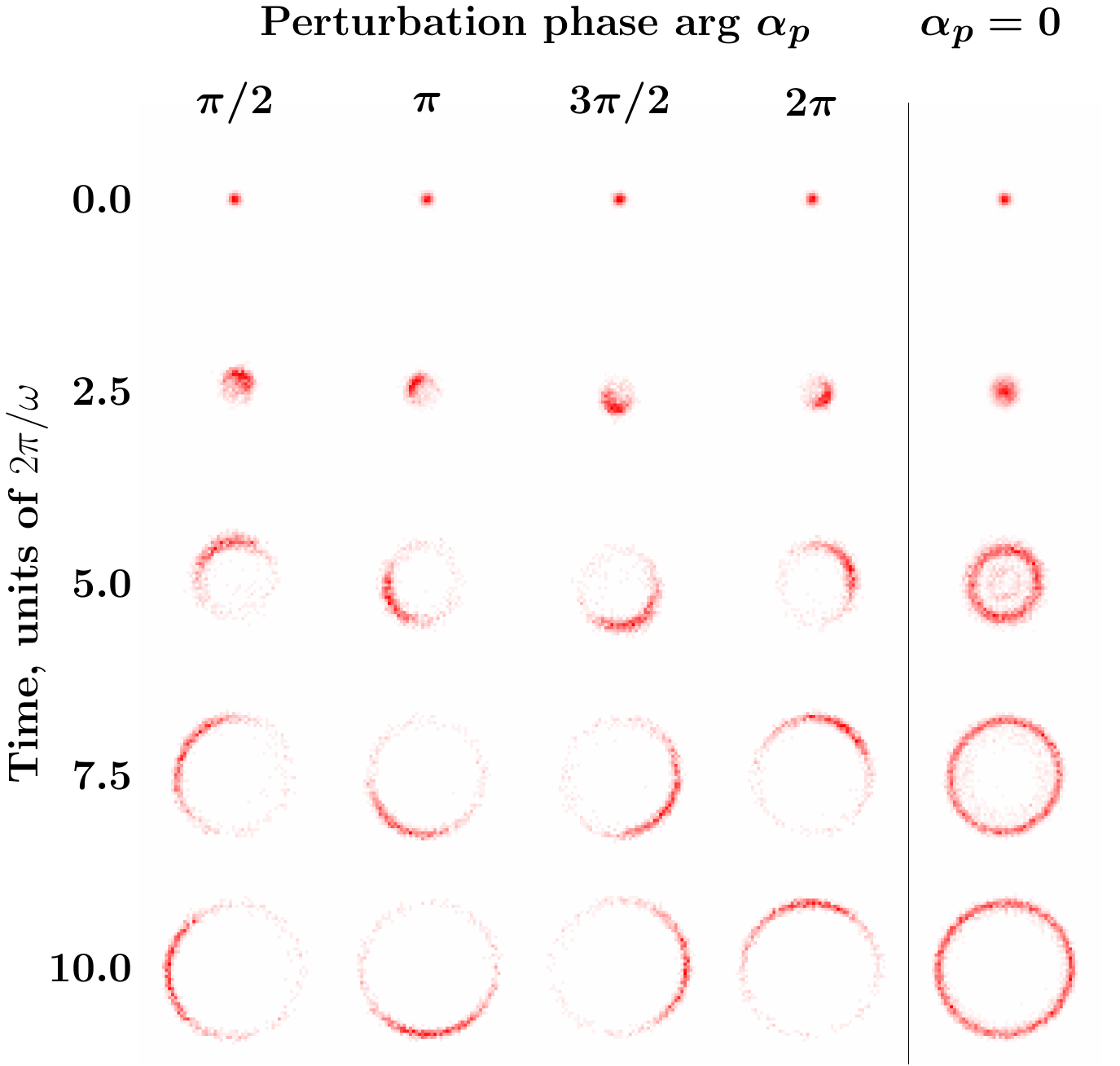}
    \caption{Evolution of the 3rd cavity mode's Husimi function depending on the perturbation's phase (left to right) and evolution time (top to bottom). For comparison, the rightmost column shows the time-evolving Husimi function for the unperturbed system.}
    \label{fig:husgrid}
\end{figure}

\begin{figure}[b]
    \centering
    \includegraphics[width=\linewidth]{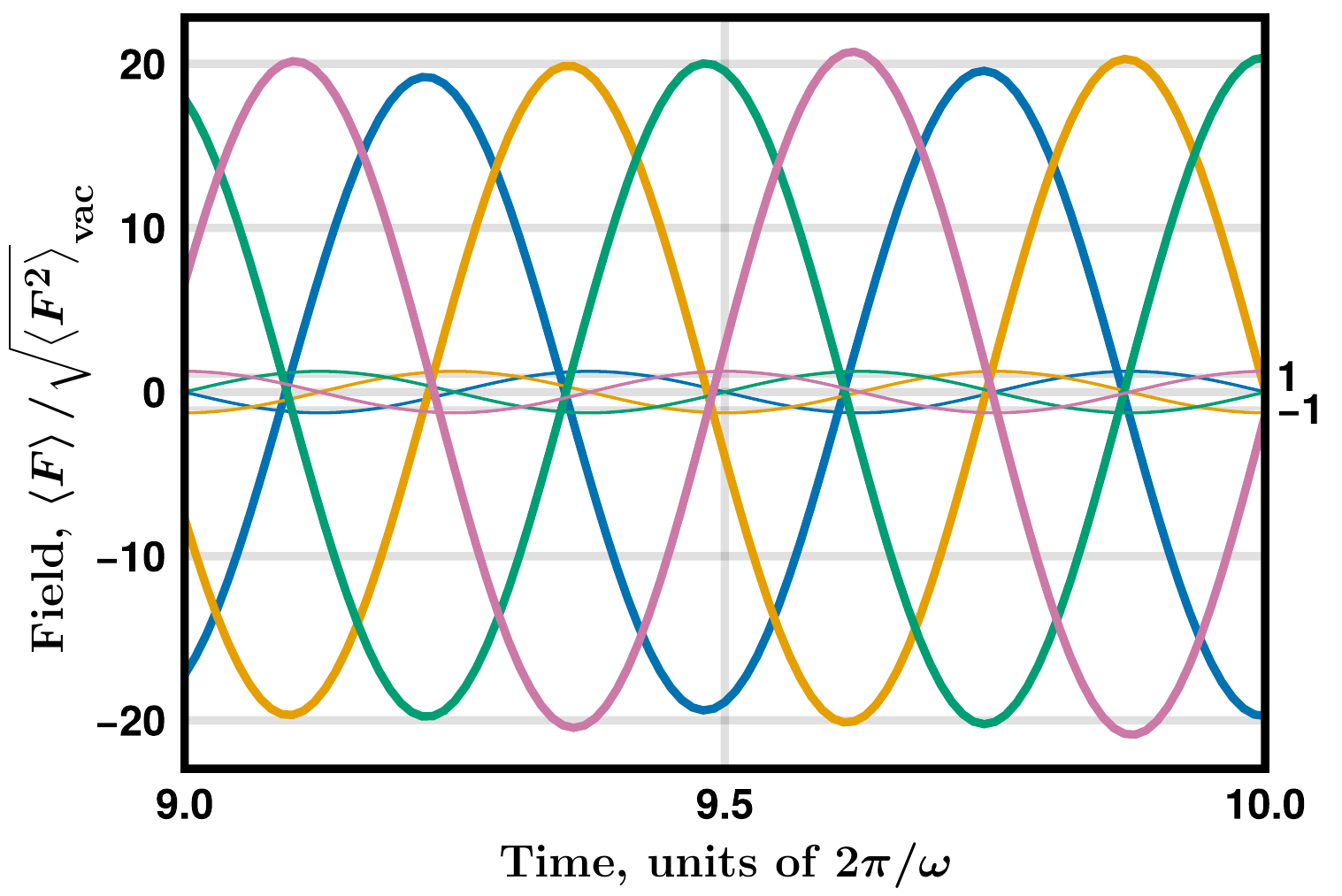}
    \caption{Time-dependent output field shapes plotted for varying perturbing field phase. The phases plotted are: $\pi/2$ (blue), $\pi$ (yellow), $3\pi/2$ (green), $2\pi$ (magenta). The thin lines indicate the initially introduced fluctuation.}
    \label{fig:fieldpert} 
\end{figure}

The cavity has a generic Hamiltonian 
\begin{equation}\label{hb-k}
   \hat{H}_B = \int d \omega \omega \ \hat{a}_\omega^\dagger \hat{a}_\omega 
\end{equation}

Its electric field operator is then, for a particular system-dependent function $c(\omega)$

\begin{equation}\label{eq:f-op}
 \hat{F} = \int d\omega c(\omega) (\hat{a}_\omega+\hat{a}^\dagger_\omega)  
\end{equation}

The coefficients $c(\omega)$ incorporate both the cavity's density of states and the emitting dipole's coupling strength to the particular mode frequency. In practice, it is sometimes useful to define $c(\omega)$ parametrically, making a change of variables $\omega\rightarrow k$ and defining a dispersion $\omega(k)$. In our model, where the cavity's photonic modes are considered to be similar to a narrowband waveguide, the frequency modes are described using a standard tight-binding approximation with the next-neighbor coupling constant $h$. Within this approximation, the cavity eigenstates are excitations indexed by a wavenumber $k \in [0, \pi)$:
\begin{equation}\label{omega-k}
 \omega(k) = \epsilon + 2h_\omega \cos{k} ; \ \
 c(k) = h_c \sqrt{2 \pi} \sin{k} \ \ 
 \end{equation}

The coefficients $c(k)$ are then used to evaluate the operator $\hat{F}$ as described in the Supplementary, giving us the interaction operator:

\begin{equation}\label{hi-k}
 \hat{H}_I = \hat{d} \otimes \hat{F}
 \end{equation}

Here, the operator $\hat{d} = \sum_{j=1}^N \sigma_j^x$ is the total dipole operator summed over all the atoms. 
In the calculations, we set $h_c=0.05\omega_0, h_\omega = 0.025\omega_0$. In addition, all excitations in the waveguide experience decay at a rate $\kappa = h_\omega = 0.025\omega_0$. For more details on the numerical implementation, see the Supplementary.

A head-on approach to this problem is ill-advised. An exact diagonalization solution of the time-dependent Schrödinger equation would be prohibitively expensive due to the fact a continuously driven system emits photons continuously, which in its turn causes the required memory space to also scale up continuously. Given the broadband setup we have chosen, storing a quantum state beyond 5-10 total emitted photons becomes infeasible. For this reason, we use instead the Hierarchy of Pure States (HOPS) method~\cite{hartmann2017exact, polyakov2019dressed}. Unlike exact diagonalization, HOPS treats the joint state of the emitters and the light field as a sequence indexed by a number $\xi$, consisting of classical amplitudes $\bm{\alpha}_\xi$ and their respective emitter state projectors $\ketbra{\psi(\bm{\alpha}_\xi)}$. The $\bm{\alpha}_\xi$ have the physical meaning of coherent states measured by an external detector. Mathematically, they are draws from the light field's multimode Husimi function $Q(\bm{\alpha})$. By being c-numbers, $\bm{\alpha}$, and the respective photon counts, can be scaled arbitrarily high without incurring any costs on memory space. 

The choice to center the cavity spectrum at the second harmonic of the driver is due to the following considerations. 
On the one hand, individual atoms driven by a classical field and preserving the inversion and time shift symmetry cannot emit the second harmonic. Indeed, the uncorrelated second harmonic emission by individual atoms remains negligible in our case.

On the other hand, many systems with a large amount of degrees of freedom tend toward an equilibrium. In our case, however, the atoms can only reach such equilibrium with the electromagnetic field when their gain arising from photon absorption from the driving field evens out with the loss happening due to the emission. Given that the individual atom's second harmonic emission amplitude is negligible, reaching an equilibrium becomes impossible for uncorrelated atoms. This way, we force the system to undergo a collective spontaneous symmetry-breaking transition into a correlated state where the second harmonic can be emitted efficiently.

In Fig.~\ref{fig:husgrid}, we plot the Husimi functions of a single discretized mode within the cavity's frequency spectrum depending on the complex amplitude $\alpha_p$. In the initial vacuum Husimi function, the fluctuations $\bm{\alpha}_\xi$ follow the Gaussian probability distribution. As the system begins to evolve under the action of the $\omega$ laser field, the $2\omega$ vacuum fluctuations unlock emission at the second harmonic frequency, creating a dipole which is phase-locked to these fluctuations. As discussed in \cite{klimkin2025spontaneous}, these fluctuations are entangled with the atomic dipole, causing an initially shot-noise fluctuation $\alpha$ to give rise to emission vastly exceeding $\alpha$ in magnitude while still reproducing it. As such, for a purely vacuum initial state characterized by a phase-independent Gaussian Husimi function, shown in the rightmost column of Fig.~\ref{fig:husgrid}, the emitted field's Husimi function is phase-independent likewise.

However, the addition of an initial classical field $\alpha_p$ imbalances the vacuum fluctuations, causing the resulting propagated Husimi function to develop a skew in one particular quadrature direction instead of remaining phase-independent. In the course of laser-driven evolution, as Fig.~\ref{fig:husgrid} shows, the ring-like shapes of the quantum mode's Husimi function become lopsided for a nonzero $\alpha_p$. As the phase of the perturbing field $\alpha_p$ is varied, we observe that the shape of the emitted field's Husimi function varies concurrently, a $\pi/2$ change in the phase of $\alpha_p$ bringing about a $\pi/2$ rotation of the Husimi Q. %, remaining phase-locked to the perturbation.

As demonstrated in Fig.~\ref{fig:fieldpert}, the described Husimi function skewing translates readily to a drastic change in field observables. We plot with respect to time the average fields $\expval{\hat{F}}$ calculated for every value of $\alpha_p$, with $\expval{\hat{F}}$ defined in (\ref{eq:f-op}). The fields plotted are normalized to the average shot noise amplitude, $\sqrt{\expval{\hat{F}^2}{\text{vac}}}$. 

As seen in Fig.~\ref{fig:fieldpert}, in spite of the fact that the input coherent field remains below the average shot noise amplitude at all times, the output fields rise drastically above the shot-noise level, allowing for phase detection by a macroscopic device affected by a classical noise. Remarkably, these amplified fields retain the phase information imparted on them by the microscopic initial field. As such, a $\pi/2$ phase delay in the initial field translates to an equivalent $\pi/2$ phase delay in the output. 

It is the conversion of classical radiation into second-harmonic emission tightly entangled with the atomic state which allows for the powerful quantum amplification described in this work. An initially small vacuum fluctuation triggers an emission avalanche, causing the atoms to develop a dipole oscillating at a parity-prohibited frequency, with this oscillation being phase-locked to the original vacuum fluctuation. In contrast to regular avalanche detection, in which the time of arrival of a particle is transferred to the time a current spike is registered, the phase of the fluctuating quantum field is carried over to the phase of the observable macroscopic field's fluctuations \cite{klimkin2025spontaneous}. 

However, our present work demonstrates that the macroscopic output field's averages, as opposed to fluctuations, can as well be used to detect microscopic deviations of the input field from the vacuum. Indeed, the $\alpha_p$ shift serves as a bias to the vacuum fluctuations, redistributing Husimi Q probability density between phase angles. In Fig.~\ref{fig:husgrid}, every point sampled from the ring-like Husimi function corresponds to an emission amplitude far exceeding the shot noise level, resulting from an emission avalanche in its own quadrature direction. As such, our concept of quantum sensing involves an \textit{imbalancing} of these emission avalanches that would otherwise cancel out, causing the resulting imbalanced average to greatly surpass the original imbalancing force.

Our method is generalizable to arbitrary mode setups. The key principle of our work is the fact that emission into modes prohibited by dynamical symmetries can become self-amplifying. In our work, we use the simplest example of such a dynamical symmetry, given by the fact that a monochromatic light wave inverts its sign after half its period, which prohibits second harmonic emission in central symmetric media. However, a far broader range of such symmetries is known in general~\cite{neufeld2019floquet}, prompting modified setups exploiting e.g. time-reversal and rotation symmetries. 

One platform warrants our particular attention as a potential proving ground.
An ensemble of alkaline atoms trapped within an optical cavity such that their regular emission is suppressed by the Purcell effect can be excited with a moderately intense ($10^{11} - 10^{12} \text{ W/cm}^2$) laser source. These can be expected to undergo a Rabi oscillation, begin exchanging virtual photons, transition to a collective entangled state, and emit radiation bursts in a cooperative way, as predicted by this work. The relatively weak coupling to cavity modes can be offset by the potentially very large number of atoms involved, even at modest gas pressures.

Thanks to the new addition to the toolset applied to the problem of quantum harmonics generation, our work shows how non-perturbative interactions can reveal the fluctuations of a quantum state of light with an arbitrary magnification. However, not only the measurement, but also the manipulation of quantum states of light is held back by the insufficiency of computational methods used to describe their non-perturbative interaction with matter. We believe our work bridges this gap.
\smallskip

\begin{acknowledgments}
%We acknowledge , 

N.K. acknowledges support of CRC 1477 "Light-Matter Interactions at Interfaces", Project No. 441234705. M.I. acknowledges support of ANR-DFG project ”Generation of bright non-classical light based on high harmonics and its use in quantum spectroscopy,” project No. 545591821, and ISF-DFG project ”Quantum optics of high harmonic generation in resonant media”, project No. 560535838.
N.K. acknowledges Evgeny A. Polyakov's clarifications of Ref.~\cite{polyakov2019dressed}, Vladislav Sukharnikov and Stasis Chuchurka's advice on the HOPS method, and useful advice from Johannes Feist. M.I. and N.K. acknowledge many enlightening discussions with Alexey Gorlach, Ido Kaminer, Moti Segev, and Stefanos Carlström. In addition, N.K. acknowledges Moti Segev's help in maintaining and continuing this work by assisting his reaching shelter during an inbound missile alert. 
\end{acknowledgments}

\nocite{klimkin_2026_20056341}
\nocite{klimkin2025codebase}

\bibliography{refs}

\appendix
\section{\label{app:numset}Numerical setup}
The non-Markovian method we adopt is a modification of the Hierarchy of Pure States (HOPS) \cite{hartmann2017exact}. In addition to giving us access to field observables, this method, as argued in \cite{polyakov2019dressed}, is exceptionally well-suited for dynamically-driven systems. We implement the HOPS method as described in \cite{polyakov2019dressed}. Our implementation allows us to investigate the mutual interaction of an arbitrarily large number of emitters with a narrowband electromagnetic vacuum. We obtain the Husimi function of the generated light and then quantify its properties by computing its antinormally-ordered quantum correlators.

Our joint system includes a small reduced system (S), a large bath of non-interacting electromagnetic modes (B), and an interaction term connecting them:
\begin{equation}\label{eq:hgen}
 \hat{H} = \hat{H}_S + \hat{H}_I + \hat{H}_B 
 \end{equation}
We model the photonic bath (B) using a tight-binding-like Hamiltonian for a semi-infinite chain of bosonic states localised on evenly spaced sites numbered by $j$, connected to one another by a nearest-neighbor hopping. The Hamiltonian is:
\begin{equation}\label{eq:tbbh}
 \hat{H}_B = \epsilon \sum_{j=1}^\infty \hat{a}^\dagger_j \hat{a}_j + h \sum_{j=1}^\infty \hat{a}^\dagger_j \hat{a}_{j+1} + \text{h.c.} 
 \end{equation}
This photonic chain interacts with a reduced system (S) made up by an array of $N_e$ atoms with $n$ orbitals. As elaborated in \cite{pizzi2023light}, its Hamiltonian can be written in terms of effective bosonic creation and annihilation operators $\hat{A}^\dagger_m$, $\hat{A}_m$, with $\left[\hat{A}_m, \hat{A}^\dagger_{m'}\right] = \delta_{m m'}$, as:
\begin{equation}\label{ham-s}
 \hat{H}_S = \sum_{m=1}^n \epsilon_m \hat{A}^\dagger_m \hat{A}_m + F(t) \sum_{m=1}^n \sum_{m'=1}^n d_{m m'} \hat{A}^\dagger_m \hat{A}_{m'} 
 \end{equation}
The Hilbert space for this Hamiltonian corresponds to $n$ effective bosonic modes filled by a total of $N_e$ atoms. 

If each emitter has $n=2$ and its  basis is chosen such that $\epsilon = (\omega_0/2, -\omega_0/2)$, $\hat{d} = d_0 \hat{\sigma}_x$, the total Hamiltonian can be written in a familiar form:
\begin{equation}\label{hs-S}
 \hat{H}_S = (\omega_0/2) \hat{S}_z + d_0 F(t) \hat{S}_x 
 \end{equation}
where $\hat{S}_x$, $\hat{S}_z$ belong to a set of operators for describing the collective pseudospin of the emitters:

\begin{eqnarray}
    \hat{S}_+ &:=& \hat{A}_2^\dagger \hat{A}_1\nonumber\\
    \hat{S}_- &:=& \hat{A}_1^\dagger \hat{A}_2\nonumber\\
    \hat{d} \equiv \hat{S}_x &:=& \hat{S}_+ + \hat{S}_- \\
    \hat{S}_y &:=& i (\hat{S}_+ - \hat{S}_-) \nonumber\\ 
    \hat{S}_z &:=& \hat{A}_1^\dagger \hat{A}_1 -\hat{A}_2^\dagger \hat{A}_2\nonumber
\end{eqnarray}

The interaction (I) operator is 
\begin{equation}
 \hat{H}_I = h \hat{d} (\hat{a}_1 + \hat{a}_1^\dagger) 
 \end{equation}
 
A generic photonic bath Hamiltonian is given in terms of the continuous-frequency creation and annihilation operators $\hat{a}^\dagger(\omega)$, $\hat{a}(\omega)$ following the conventional commutation relations:

\begin{equation}
    \left[\hat{a}(\omega), \hat{a}^\dagger(\omega')\right] = \delta(\omega - \omega')
\end{equation}

\begin{equation}\label{hb-omega}
 H_B = \int dk \omega \hat{a}^\dagger (\omega) \hat{a}(\omega)
\end{equation}

The Hamiltonian (\ref{hb-omega}) does not include a density of states, and does not account for the bath being spectrally limited. Instead, both the density of photonic states and the frequency-dependent photon mode coupling are described by the generic frequency-dependent coupling coefficient $c(\omega)$ in the integral below:

\begin{equation}\label{hi-omega}
 \hat{H}_I = \hat{d} \left(\int d\omega c(\omega) \hat{a}(\omega) + \text{h.c.} \right)
 \end{equation}

For a photonic energy band modeled in the tight-binding approximation as given by (\ref{eq:tbbh}), we Fourier transform along the semi-infinite $m$ dimension to arrive at non-interacting delocalized modes indexed by the wavenumber $k \in [0, \pi)$. Instead of $c(\omega)$, we then work with a $c(k)$ and $\omega(k)$ defined in terms of $k$ as a parameter. For $\epsilon$, $h$ matching the ones used in (\ref{eq:tbbh}), their exact form is:
 
\begin{equation}%\label{omega-ks}
    \omega(k) = \epsilon + 2h \cos{k} 
\end{equation}
\begin{equation}%\label{c-k}
    c(k) = h \sqrt{2 \pi} \sin{k} 
\end{equation}

To be useful for a practical simulation, the mode continuum needs to be discretized into $N_m$ modes, numbered by an index $\nu = 1\ldots N_m$. We choose the modes such that the grid over $k$ is uniform:

\begin{equation}
    k_\nu = \frac{\pi \nu}{N_m + 1}
\end{equation}

We then set $\omega_\nu = \omega(k_\nu), c_\nu = c(k_\nu)$. The Hilbert space for quantum photons is then truncated at $N_p$ total photons -- i.e. we include every quantum photon state $\ket{n_1, \ldots, n_{N_m}}$ for $\sum_{\nu=1}^{N_m} n_\nu \leq N_p$. Then, labeling the cavity-side operator in (\ref{hi-omega}) as $\hat{b}$:

\begin{equation}
    \hat{b} := \int d\omega c(\omega) \hat{a}(\omega)
\end{equation}

In order to preserve commutation relations between the discretized creation and annihilation operators $\left[\hat{a}_\nu, \hat{a}^\dagger_{\nu'}\right] = \delta_{\nu\nu'}$, we discretize the operator $\hat{b}$ as:

\begin{equation}\label{eq:bd}
    \hat{b} = \sum_{\nu=1}^{N_m} c_\nu \hat{a}_\nu \sqrt{\Delta \omega_\nu}
\end{equation}

Respectively, the discretized field Hamiltonian is:

\begin{equation}\label{eq:hd}
    \hat{H}_B = \sum_{\nu=1}^{N_m} \omega_\nu \hat{a}_\nu^\dagger \hat{a}_\nu
\end{equation}

\section{\label{app:calcobs}Calculating observables}

By being stochastic, our method yields the overall system-bath state in the form of many statistical samples numbered by an index $\xi$ running from $1$ to the total number of samples $N_\text{batch}$. Each of these samples contains a wavefunction $\ket{\Psi_\xi}$ and a draw from the bath's Husimi distribution $Q(\bm\alpha)$, designated as $\bm\alpha_\xi$, correlated with $\ket{\Psi_\xi}$. The wavefunction $\ket{\Psi_\xi}$ will then be called conditional, or equivalently, conditioned on $\bm{\alpha}_\xi$. 

The Hilbert space of the computational wavefunctions $\ket{\Psi_\xi}$ includes both the emitters and quantum photons, called ''virtual'', in contrast to the ''real'' photons corresponding to the $\bm{\alpha}_\xi$. To arrive at the partial density matrix of the emitters, we average the conditional projectors, themselves projected onto a quantum vacuum of the virtual photons. That is to say, the reason the virtual photons are called virtual is that their presence is not directly observable by a distant classical detector.

\begin{equation}
    \hat{\rho}_a := \frac{1}{N_\text{batch}} \sum_\xi \frac{\bra{\boldsymbol{0}}\cdot\ketbra{\Psi_\xi}\cdot \ket{\boldsymbol{0}}}{\left|\braket{\Psi_\xi}{\boldsymbol{0}}\right|^2}
\end{equation}

This density matrix can now be used to calculate the expectation value of every emitter-side operator $\hat{O}$:

\begin{equation}
\expval{O}_\text{stoch} := \text{tr}\left(\hat{O} \hat{\rho}_a\right) \approx \expval{\hat{O}}
\end{equation}

The observables of the photonic state are defined by the classical statistics of $\bm{\alpha}$. As such, for an arbitrary antinormally-ordered quantum average characterized by orders $m_1, m_2, ...$, $n_1, n_2, ...$ for modes $\mu_1, \mu_2, ...$, $\nu_1, \nu_2, ...$, there's a stochastic average which converges to the correct quantum value:
\begin{equation}\label{stoch-avg}
 \frac{1}{N_\text{batch}} \sum_\xi \left(\prod_l \alpha_{\xi; \mu_l}^{m_l} \prod_l (\alpha_{\xi; \nu_l}^{n_l})^*\right) \approx \expval{\prod_l \hat{a}_{\mu_l}^{m_l} \prod_l (\hat{a}_{\nu_l}^{n_l})^\dagger}
 \end{equation}
 
This fact allows us to recover photonic observable averages. Of particular interest to us are the average fields calculated as the following stochastic expectation value: 

\begin{equation} 
    \expval{\hat{F}} = \frac{1}{N_\text{batch}} \sum_\xi \left(f_\xi + f_\xi^*\right)
\end{equation}

\begin{equation}\label{eq:f-xi}
    f_\xi := \sum_{\nu=1}^{N_m} c_\nu \alpha_{\xi; \nu} \sqrt{\Delta \omega_\nu}
\end{equation}

\section{Numerical implementation}

The resulting TDSE is a system of coupled nonlinear ordinary differential equations which amounts to a slight rewriting of the ones given in \cite{polyakov2019dressed} for purposes of easier computational treatment. The basic Hamiltonian coincides with the one given in (\ref{eq:hgen}). However, it is supplemented by an additional non-Hermitian term describing interaction with a stochastically-sampled vacuum fluctuation $\bm{\alpha}$, as well as a Hermitian term describing an initially existent displacement of the Husimi function. Following \cite{polyakov2019dressed}, who described its application to describing thermal photonic states, we will call this initial displacement $\bm{\alpha}_{th}$. The advantage of this scheme over exact diagonalization is that in many cases~\cite{hartmann2017exact, polyakov2019dressed}, this additional classical interaction reduces the number of quantum photons involved in the resultant evolution, shrinking drastically the dimension of their Hilbert space.
\begin{equation}
 \hat{H}_0 (t) = \hat{H}_S (t) + \hat{H}_I + \hat{H}_B 
\end{equation}
\begin{align}\label{ham-stoch}
 \hat{H} [\bm{\alpha}, \ket{\psi}] (t) =& \hat{H}_0 (t) +  \hat{d} \sum_{\nu = 1}^{N_m} c_\nu^* \alpha_\nu^* \sqrt{\Delta \omega_\nu} - \expval{\hat{d}}^*_\psi \hat{b} \nonumber \\ & + \hat{d} \sum_{\mu = 1}^{N_{th}} \left(c_{\mu}^* \alpha_{th;\mu}^* + c_{\mu} \alpha_{th;\mu}\right) \sqrt{\Delta\omega_{\mu}}
\end{align}

Here, the definition of $\hat{b}$ follows (\ref{eq:bd}). In this numerical discretization procedure, $N_{th}$ does not have to be equal to $N_m$. In this case, $c_\mu$ is discretized for $N=N_{th}$ according to (\ref{eq:bd}). Specifically, in our case $N_{th} = 1$, corresponding to a continuous monochromatic excitation initially present in the waveguide. Then, the 

The equations of evolution for $\alpha_{th}$ follow the ones written for Husimi function fluctuations $\alpha$, except that $\alpha_{th}$ do not have a source term.

Following \cite{muller2025quantum,muller2025genuine}, we also add a decay term $\kappa$ acting as an artificial cutoff on the photonic modes' memory function. As such, in the Hamiltonian \ref{eq:hd}, the frequencies $\omega_\nu \rightarrow \omega_\nu - i\kappa$. Likewise, we separate the vacuum fluctuations $\alpha_{\nu; \xi}$ into two parts. One is a sourceless part $\alpha_0$ that corresponds to the random draw from the Husimi function, does not decay, and is not affected by a source term. Any excitations created are described by the shift $\delta \alpha$ representing the decaying memory function. Then $\alpha = \alpha_0 + \delta \alpha$.

With that said, the final equations we use are:

\begin{equation}\label{syseq}
    \begin{cases}
        i \partial_t \ket{\psi_\xi (t)} &= \hat{H}[\bm{\alpha}_\xi (t), \ket{\psi_\xi (t)}] (t) \ket{\psi_\xi (t)}\\
  i \partial_t \alpha_{0; \nu, \xi}(t) &= \omega_\nu \alpha_{0; \nu, \xi}(t)\\
  i \partial_t \delta \alpha_{\nu, \xi}(t) &= (\omega_\nu - i \kappa) \alpha_{\nu; \xi}(t) + c_\nu^* \expval{\hat{d}}_{\psi_\xi}\\
  i \partial_t \alpha_{th; \nu, \xi}(t) &= \omega_\nu \alpha_{th; \nu, \xi}(t)\\
    \end{cases}
\end{equation}

The absence of a $+\text{h.c.}$ in (\ref{ham-stoch}) is not in error. As opposed to describing the wavefunction of the full system $\ket{\Psi}$ like TDSE normally does, (\ref{syseq}) deals with a conditional wavefunction for the reduced system $\ket{\psi_\xi(t)}$, measured in coincidence with its coherent state $\bm\alpha_\xi$. As such, its evolution is non-unitary, and the resulting Hamiltonian non-Hermitian. Consequently, the resulting stochastic wavefunctions $\ket{\psi_\xi}$ are also not normalized. Expectation values in the form $\expval{O}_\psi$ must be understood as normalized averages over vacuum-projected computational wavefunctions:
\begin{equation}
 \expval{\hat O}_\psi \equiv \frac{\braket{\psi}{\bm{0}} \hat{O} \braket{\bm{0}}{\psi}}{|\braket{\psi}{\bm{0}}|^2} %\frac{\expval{\hat O}{\psi}}{\braket{\psi}}
\end{equation}

We solve the equations given by (\ref{syseq}) by starting at the initial conditions:

\begin{equation}
    \begin{cases}
        \ket{\psi_\xi (-\infty}) &= \ket{g}\\
        \bm{\alpha}_{0; \xi} (-\infty) &\sim \mathcal{C N}(\bm{0}, \bm{1})\\
        \delta \bm{\alpha}_{0; \xi} (-\infty) &= 0\\
        \alpha_{th; \xi} (-\infty) &= \alpha_p\\
    \end{cases}
\end{equation}

For streamlining purposes, the solutions for the different initial conditions are stored as multidimensional tensors and solved jointly. The solution-dependent coefficients proportional to e.g. $\alpha_{\nu; \xi}$ and $\expval{\hat{d}}_{\psi_\xi}$ are applied after matrix multiplication. 

(\ref{syseq}) is a system of nonlinear ordinary differential equations for $\ket{\psi}$ and $\bm{\alpha}$ which can still be linearised in a straightforward way. One commonly used approach to solving problems of this kind are exponential Rosenbrock-type methods \cite{caliari2009implementation, hochbruck2010exponential}. The selected solver algorithm is the Rosenbrock-Euler scheme. The $\varphi$-functions required for the Rosenbrock method's operation are computed as Taylor expansions up to the floating point error. All calculations are done in single (FP32) floating-point precision. The error tolerance is set to $10^{-4}$. For these parameters, a photonic band with $N_p=10$ virtual photons and $N_m = 4$ discretized modes, as well as $N_e=60$ atoms as per the main text, the runtime of a solution encompassing 4096 trajectories we attain on a single NVIDIA A100 GPU is approximately 12 hours.

\end{document}